\documentclass[%
 reprint,
superscriptaddress,
 amsmath,amssymb,
aps,
prl,
]{revtex4-1}

\usepackage{graphicx}
\usepackage{dcolumn}
\usepackage{bm}
\usepackage{gensymb}
\usepackage{floatrow}
\usepackage{hyperref}
\usepackage{xcolor}
\hypersetup{
    colorlinks=true,
    linkcolor=blue,
    citecolor=blue,
    urlcolor=blue,
}
\usepackage{color,soul}
\usepackage{adjustbox} 
\usepackage{tabularx,booktabs} 
\begin{document}
\preprint{APS/123-QED}

\title{Tuning spin-torque nano-oscillator nonlinearity using 
He$^{+}$ irradiation} 

\author{Sheng~Jiang}
\affiliation{Department of Applied Physics, School of Engineering Sciences, KTH Royal Institute of Technology, Electrum 229, SE-16440 Kista, Sweden}
\affiliation{Department of Physics, University of Gothenburg, 412 96, Gothenburg, Sweden}

\author{Roman Khymyn}
\affiliation{Department of Physics, University of Gothenburg, 412 96, Gothenburg, Sweden}

\author{Sunjae~Chung}
\affiliation
{Department of Applied Physics, School of Engineering Sciences, KTH Royal Institute of Technology, Electrum 229, SE-16440 Kista, Sweden}
\affiliation{Department of Physics and Astronomy, Uppsala University, 751 20 Uppsala, Sweden}

\author{Quang~Tuan~Le}
\affiliation{Department of Applied Physics, School of Engineering Sciences, KTH Royal Institute of Technology, Electrum 229, SE-16440 Kista, Sweden}
\affiliation{Department of Physics, University of Gothenburg, 412 96, Gothenburg, Sweden}

\author{Liza Herrera Diez}
\affiliation
{Institut d'Electronique Fondamentale, CNRS, Universit\'e Paris-Sud, Universit\'e Paris-Saclay, 91405 Orsay, France}

\author{Afshin~Houshang}
\affiliation{Department of Physics, University of Gothenburg, 412 96, Gothenburg, Sweden}
\affiliation{NanOsc AB, Kista 164 40, Sweden}

\author{Mohammad Zahedinejad}
\affiliation{Department of Physics, University of Gothenburg, 412 96, Gothenburg, Sweden}

\author{Dafin\'e Ravelosona}
\affiliation
{Institut d'Electronique Fondamentale, CNRS, Universit\'e Paris-Sud, Universit\'e Paris-Saclay, 91405 Orsay, France}
\affiliation
{Spin-Ion Technologies, 28 rue du G\'en\'eral Leclerc, 78000 Versailles Cedex, France}

\author{Johan~\AA{}kerman}
 \altaffiliation[Author to ]{johan.akerman@physics.gu.se}
\affiliation
{Department of Applied Physics, School of Engineering Sciences, KTH Royal Institute of Technology, Electrum 229, SE-16440 Kista, Sweden}
\affiliation
{Department of Physics, University of Gothenburg, 412 96, Gothenburg, Sweden}
\affiliation{NanOsc AB, Kista 164 40, Sweden}
\email{johan.akerman@physics.gu.se}

\date{\today}

\begin{abstract}
We use He$^+$ irradiation to tune the nonlinearity, $\mathcal{N}$, of all-perpendicular spin-torque nano-oscillators (STNOs) using  the He$^+$ fluence-dependent perpendicular magnetic anisotropy (PMA) of the [Co/Ni] free layer. Employing fluences from 6 to 20$\times10^{14}$~He$^{+}$/cm$^{2}$, we are able to tune $\mathcal{N}$ in an in-plane field from strongly positive to moderately negative. As the STNO microwave signal properties are mainly governed by $\mathcal{N}$, we can in this way directly control the threshold current, the current tunability of the frequency, and the STNO linewidth. In particular, we can dramatically improve the latter by more than two orders of magnitude. Our results are in good agreement with the theory for nonlinear auto-oscillators, confirm  theoretical predictions of the role of nonlinearity, and demonstrate a straightforward path towards improving the microwave properties of STNOs.

\begin{description}
\item[DOI]
XXXXXX.
\end{description}
\end{abstract}

\pacs{Valid PACS appear here}
       
\maketitle


Spin-torque nano-oscillators (STNOs) are among the most promising candidates for nanoscale broadband microwave generators~\cite{Deac2008,Maehara2013a,Mohseni2013,Maehara2014,Chen2016,Banuazizi2017a} and detectors~\cite{Braganca2010b,Miwa2014,Fang2016}. STNOs can generate broadband microwave frequencies ranging from hundreds of MHz to the  sub-THz~\cite{Pribiag2007a,Hoefer2005,Stamps2014},  controlled by both magnetic fields and dc currents~\cite{Dumas2013a,Chen2016}. Moreover, the device size can be reduced to a few tens of nanometers, which is of great opportunity for industrial applications. They can also host a range of novel magnetodynamical spin wave modes, such as propagating spin waves of different orders \cite{Bonetti2010b, Houshang2018b}, and magnetodynamical solitons, such as spin wave bullets\cite{Bonetti2010b} and droplets\cite{Mohseni2013}. 

However, the applicability of these devices has suffered from their low power emission and large linewidth. 
Nonlinear auto-oscillator theory~\cite{Kim2008,Kim2008a,Slavin2008,Andrei2009} 
explains the large linewidth 
as a result of the strong nonlinearity $\mathcal{N}$, \emph{i.e.}~the dependence of the microwave frequency on its precession amplitude. 
$\mathcal{N}$ 
can be controlled not only by the measurement conditions~\cite{Pufall2006a,Rippard2006,Gerhart2007,Consolo2008,Muduli2010a,Dumas2013a}, such as the magnitude and direction of the magnetic field, but also by the magnetic properties of the free layer of the STNO, such as the magnetic anisotropy and the effective magnetization~\cite{Andrei2009}. For instance, in an easy-plane free layer, $\mathcal{N}$ changes gradually from positive to negative values as the direction of magnetic field rotates from out-of-plane to in-plane~\cite{Kim2008,Andrei2009}. Experimental studies have corroborated \cite{Rippard2006,Kim2008,Bonetti2010b,Bonetti2012,Lee2013a,Dumas2013a,Mohseni2018b} this theoretical prediction, 
as the linewidth shows a minimum when $\mathcal{N}$ crosses zero at the critical field angle. This suggests a way to improve the linewidth by selectively reducing the nonlinearity.

Whereas all previous studies aimed at minimizing the nonlinearity have focused only on the effects of the external conditions in single devices, a more general and practical solution should be based on the intrinsic magnetic properties of the device itself. In our work, we therefore study systematically how $\mathcal{N}$ is affected by the strength of perpendicular magnetic anisotropy (PMA) $H_\mathrm{k}$ in a set of nanocontact (NC) STNOs. We show how $\mathcal{N}$ can be continuously tuned as $H_\mathrm{k}$ is controlled by He$^+$ irradiation fluence \cite{Chappert1998, Fassbender2004, Beaujour2009, HerreraDiez2015} in otherwise identical devices. Most importantly, the linewidth is dramatically improved at moderate $H_\mathrm{k}$ values, where $\mathcal{N}\rightarrow 0$. Finally, we show excellent agreement of our experimental results with nonlinear auto-oscillator theory~\cite{Andrei2009}.


The STNO devices were fabricated from all-perpendicular (all-PMA) [Co/Pd]/Cu/[Co/Ni] \cite{Chung2018PRL, Jiang2018a} and  orthogonal [Co/Pd]/Cu/NiFe spin valves (SVs). The full stack consists of a Ta (5)/Cu (15)/Ta (5)/Pd (3) seed layer, an all-PMA [Co~(0.5)/Pd~(1.0)]${\times5}$/Cu (7)/[Co~(0.3)/Ni~(0.9)]${\times4}$/Co~(0.3) or orthogonal [Co~(0.5)/Pd~(1.0)]${\times5}$/Cu~(7)/Ni$_{80}$Fe$_{20}$ (4.5) SV with a Cu(3)/Pd(3) capping layer, sputtered onto a thermally oxidized 100 mm Si wafer (numbers in parentheses are layer thicknesses in nanometers). The deposited stacks were first patterned into 8~$\mu$m~$\times$~20~$\mu$m mesas using photolithography and ion-milling etching, followed by chemical vapor deposition (CVD) of an insulating 40-nm-thick SiO$_2$ film. Electron beam lithography and reactive ion etching were used to open nanocontacts (with nominal radius of $R_{\text{NC}}$ 35 nm) through the SiO$_2$ in the center of each mesa. The processed wafer was then cut into different pieces for He$^{+}$ irradiation 
with the fluence $F$ varied from 6 to 20$\times10^{14}$~He$^{+}$/cm$^{2}$ ~\cite{Jiang2018a}. Fabrication was completed with lift-off lithography and deposition of a Cu~($500$~nm)/Au~($100$~nm) top electrode in a single run with all irradiated pieces. Our protocol hence ensures that all other properties, except the He$^{+}$ fluence, are identical from device to device.

We used our custom-built probe station for static and microwave characterization. A direct current $I_{\text{dc}}$ was injected into the  devices using a Keithley 6221 current source, and the dc voltage was detected using a Keithley 2182 nanovoltmeter. The magnetic field was applied in the plane of the film. The generated microwave signals from the STNO device were decoupled from the dc voltage via a bias-tee, amplified using a low-noise amplifier, and then recorded with a spectrum analyzer~\cite{Jiang2018,Jiang2018PRA}.



To accurately determine $M_{\text{eff}}$ of the [Co/Ni] free layer, spin-torque ferromagnetic resonance (ST-FMR) \cite{Okutomi2011,Liu2012c,MasoumehPRB2016,Mazraati2016apl,Zahedinejad2018} measurements were performed on the He$^+$-irradiated STNOs (see details in supplemental materials \cite{Supplement}). The fluence information and the obtained effective magnetization $\mu _0 M_\mathrm{eff}$ are presented in Table \ref{table1}. The value of  $M_{\text{eff}}$ ($H_\text{k}$) increases (decreases) as the fluence increases. Here, the NiFe free layer is used as a reference for a larger $M_{\text{eff}}$ sample. 

\begin{table}
\centering
\caption{Sample structure information and the calculated effective magnetization $\mu _0 M_\text{eff}$ of free layer ([Co/Ni] or NiFe) for various He$^+$-irradiation fluences. $\mu _0 M_\text{eff}$ are measured by ST-FMR (see the supplemental materials \cite{Supplement}). }
\label{table1}
\begin{tabular*}{0.99\textwidth}%
   {@{\extracolsep{\fill}}lll}
\hline \hline
Structure  & \begin{tabular}[c]{@{}l@{}} Fluence\\ ($\times 10^{14}$~He$^{+}$/cm$^{2}$)\end{tabular} & \begin{tabular}[c]{@{}l@{}}$\mu _0 \mathit{M_\text{eff}}$\\ (T)\end{tabular} \\ 
\hline
\rm{[Co/Pd]/Cu/[Co/Ni]}              & 0                 & -0.68      \\
\rm{[Co/Pd]/Cu/[Co/Ni]}              & 6                 & -0.44      \\
\rm{[Co/Pd]/Cu/[Co/Ni]}              & 10                & -0.14      \\
\rm{[Co/Pd]/Cu/[Co/Ni]}              & 20                & 0.03       \\
\rm{[Co/Pd]/Cu/NiFe}                 & -                 & 0.98       \\
\hline \hline
\end{tabular*}

\end{table}

\begin{figure}
\centering
\includegraphics[width=3.4in]{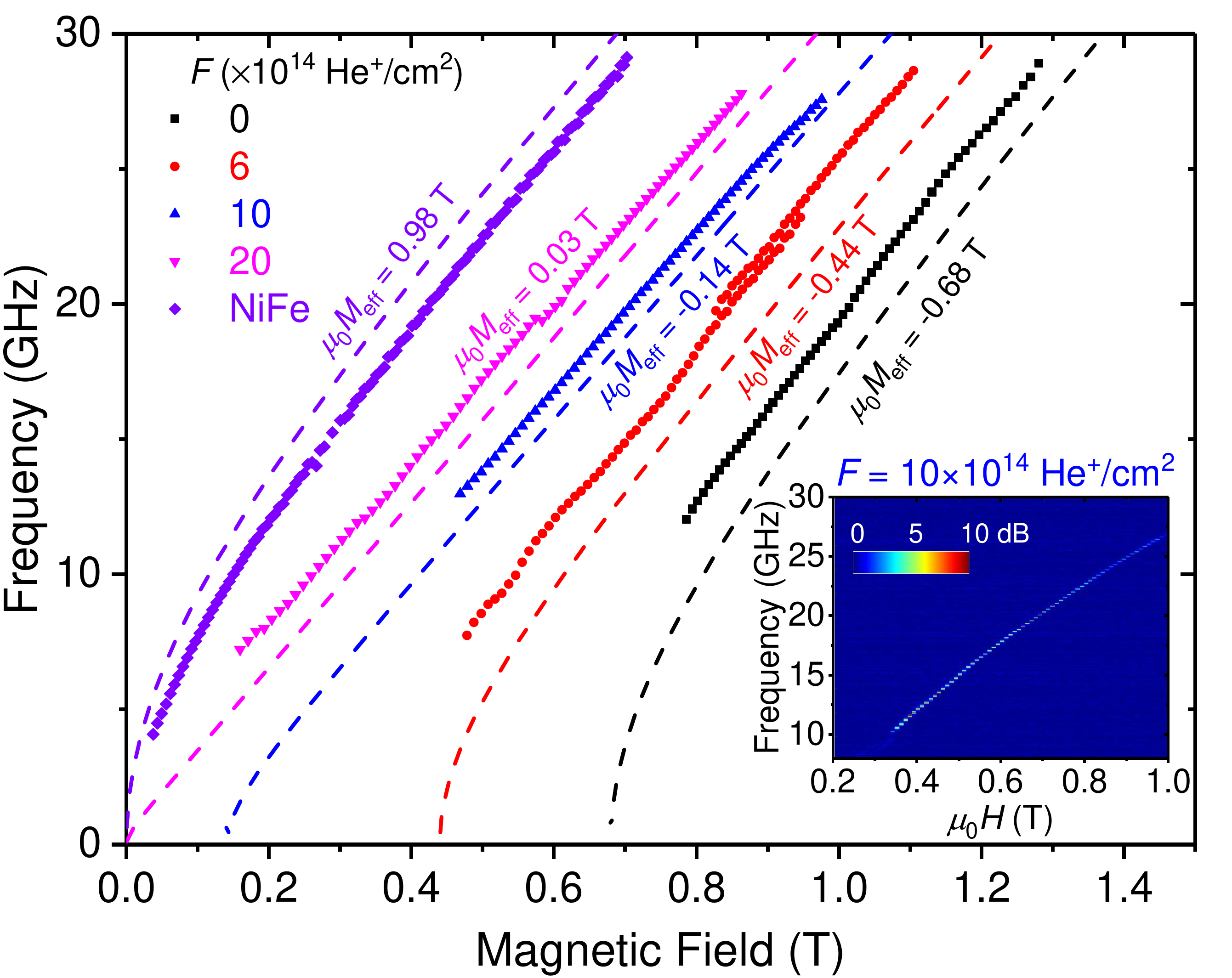}
\caption{\label{fig1} Auto-oscillation frequency versus in-plane magnetic field for various irradiated STNOs with $R_\mathrm{NC}=35$ nm. 
The dashed lines are the calculated FMR frequencies $f_{\text{FMR}}$, based on the values of $\mu_0 M_{\text{eff}}$ obtained from ST-FMR measurements \cite{Supplement}. Inset: A typical power spectral density (PSD) of an STNO with $F$ = 10$\times10^{14}$~He$^{+}$/cm$^{2}$ at $I_{\text{dc}}=-14$ mA.
}
\end{figure}

In Fig.~\ref{fig1}, we compare the calculated FMR frequency, $f_\mathrm{FMR}$, using the measured $M_{\text{eff}}$, with the  microwave signals generated from the STNO devices. 
The inset in Fig.~\ref{fig1} shows a typical power spectral density (PSD) of the microwave signals 
for a fluence of $F=10\times10^{14}$~He$^{+}$/cm$^{2}$. All PSD spectra are well fitted with a Lorentz function, and the extracted frequency $f$ versus magnetic field is presented in Fig.~\ref{fig1} with different symbols for each different fluence. All data show a quasi-linear dependence on the magnetic field, and the generated microwave frequency $f$ extends to lower values as $M_{\text{eff}}$ ($H_\text{k}$) increases (decreases). This behavior is consistent with the calculated value of the FMR frequency $f_\text{FMR}$, plotted as dashed lines in Fig.~\ref{fig1}. The overall trends of $f_\mathrm{FMR}$ are in good agreement with the auto-oscillation $f$. The difference between the calculated $f_\mathrm{FMR}$ and the measured auto-oscillation $f$ is a direct measure of the nonlinearity 
of the magnetization precession 
~\cite{Slavin2005e,Consolo2008,Bonetti2010b,Chen2016}, which is discussed in detail below. 

\begin{figure}
\centering
\includegraphics[width=3.4in]{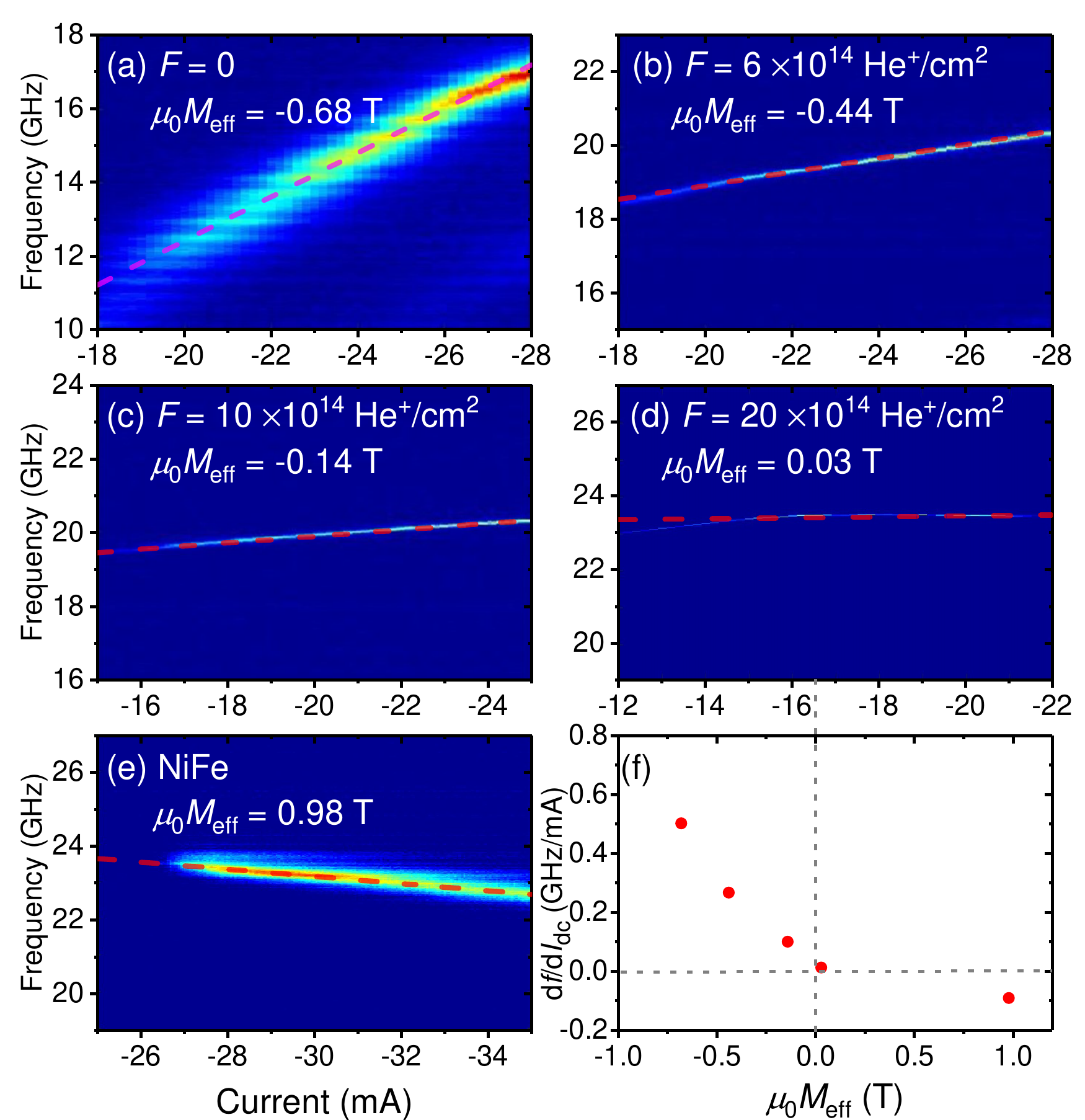}
\caption{\label{fig2} 
(a)--(e) PSD versus $I_{\text{dc}}$ in STNOs with different irradiated fluences at $\mu_0H=0.72$~T. The red dotted line represents the linear fits of the auto-oscillation frequency. (f) slope $df/dI_\mathrm{dc}$versus $\mu_0 M_{\text{eff}}$ extracted from the fits of (a)--(e). 
}
\end{figure}

We now turn to the current-induced frequency tunability. Figures~\ref{fig2}(a)--\ref{fig2}(e) show the generated microwave frequency $f$ versus dc current $I_\text{dc}$ at a fixed magnetic field, $\mu_0 H = 0.72$ T;
$f$ linearly depends on the $I_\text{dc}$ at different values of  $M_{\text{eff}}$. The current-induced frequency tunability  $df/dI_\text{dc}$ can be extracted from the slopes of linear fits which plot as each dashed line in Figs.~\ref{fig2}(a)--\ref{fig2}(e).
$df/dI_\text{dc}$ for $M_{\text{eff}}$ are then summarized in Fig.~\ref{fig2}(f). We found that \emph{i}) $df/dI_{\text{dc}}$ decreases from 0.50 GHz/mA for nonirradiated [Co/Ni] to -0.13 GHz/mA for NiFe as $M_{\text{eff}}$ increases (or $H_\text{k}$ decreases), \emph{ii}) the sign of $df/dI_{\text{dc}}$ changes from positive (for [Co/Ni]) to negative (for NiFe), consistent with the easy axis transition from out-of-plane for [Co/Ni] to in-plane for NiFe, and further details will be discussed later.

\begin{figure}
\centering
\includegraphics[width=3.4in]{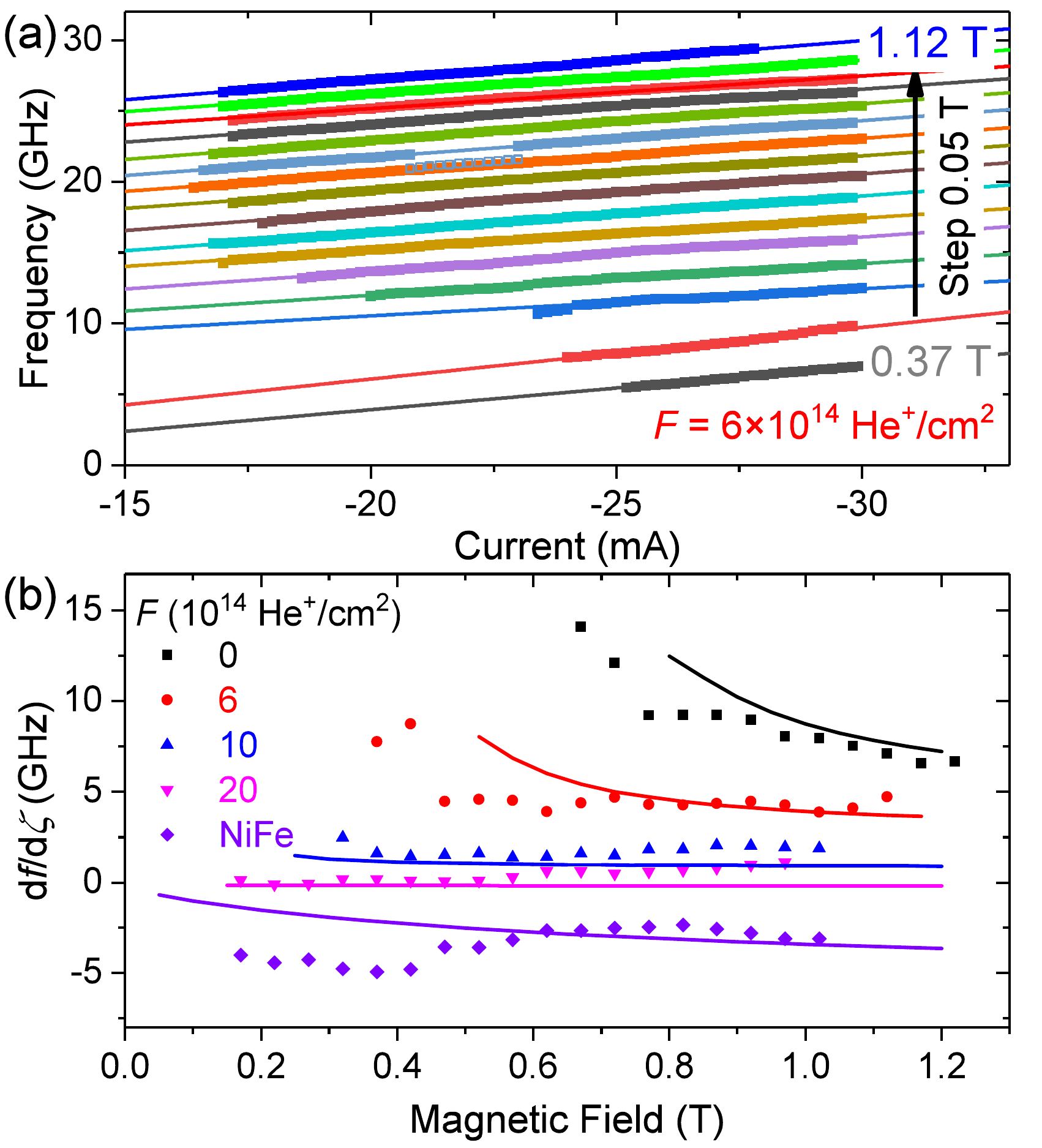}
\caption{\label{fig3} 
(a) Extracted auto-oscillation frequency $f$ vs.~$I_{\text{dc}}$ at different magnetic fields for $F$ = 6$\times10^{14}$~He$^{+}$/cm$^{2}$. Some minor frequency jumps at $\mu_0H=0.87$~T are shown as rectangular boxes, possibly due to film inhomogeneities generating different dynamical behaviors. 
(b) $df/d\zeta $ [\emph{i.e.}~$I_\text{th} (df/dI_\text{dc})$] vs.~magnetic field, where $I_\text{th}$ is extracted from the intercept of the inverse power of the auto-oscillation signals and the $df/dI_\text{dc}$ are the slopes of the linear fits of frequency as $I_\text{dc}>I_\text{th}$ (see supplemental materials \cite{Supplement}). The solid lines are the theoretical calculation from Eqs.~(\ref{eq2})--(\ref{eq4}).
}
\end{figure}

We carried out detailed measurements at different magnetic fields to understand further the behavior of $df/dI_\text{dc}$. 
Figure~\ref{fig3}(a) shows one example of extracted $f$ versus $I_\text{dc}$ at different fields, ranging from 0.37 to 1.12 T with a 0.05 T step, for $F= 6\times10^{14}$~He$^{+}$/cm$^{2}$. All data show clear linear dependencies on $I_\text{dc}$. Here we would like to define one numerical relation about the tunability, $df/d\zeta =I_\text{th} (df/dI_\text{dc})$, to compare our experimental results directly with theoretical calculation, 
where $\zeta =I_\text{dc}/I_\text{th}$ is the dimensionless supercriticality parameter ~\cite{Andrei2009} and $I_\text{th}$ is the threshold current. 
$I_\text{th}$ were extracted from plots of inverse power $1/P$ versus $I_\text{dc}$ as described in supplemental materials \cite{Supplement}. 
After obtained all $I_\text{th}$ and $df/dI_\text{dc}$ for different $M_{\text{eff}}$,  $df/d\zeta $ are represented as solid dots in Fig.~\ref{fig3}(b). All $df/d\zeta$ for different $M_{\text{eff}}$ show similar behaviors that is inverse proportional to magnetic field. It is noteworthy that the overall $df/d\zeta $ decreases as $M_{\text{eff}}$ ($H_\text{k}$) increases (decreases). It reaches around zero when the $\mu _0 M_{\text{eff}}\approx 0$ for $F= 20\times10^{14}$~He$^{+}$/cm$^{2}$.
The sign of $df/d\zeta $ for NiFe is even negative.


To understand the behavior of tunability versus $M_{\text{eff}}$ ($H_\text{k}$) from He$^+$-irradiated STNOs, we considered the nonlinear auto-oscillator theory of A. Slavin and V. Tiberkevich ~\cite{Slavin2005,Slavin2005e,Slavin2008,Andrei2009}, which was derived from universal auto-oscillation systems and has proved to be consistent with the Landau--Lifshitz--Gilbert--Slonczewski (LLGS) equation~\cite{Andrei2009}. This theory allows us to describe the experimental observation analytically. The auto-oscillation frequency $f$ generated from an STNO is expressed as:
\begin{equation}
\label{eq1}
f(I_\mathrm{dc}) =f_\mathrm{FMR}+\frac{\mathcal{N}}{2\pi}\frac{\zeta -1}{\zeta +Q},
\end{equation}
where $\mathcal{N}$ is the nonlinearity factor, $\frac{\zeta -1}{\zeta +Q}=P_0$ is the normalized power of the stationary precession, and $Q$ is the nonlinear damping coefficient. From Eq.~(\ref{eq1}), the frequency shift is mainly decided by the nonlinearity $\mathcal{N}$. Taking the derivation of Eq.~(\ref{eq1}), $df/d\zeta $ is derived as:
\begin{equation}
\label{eq2}
\frac{df}{d\zeta}=I_\text{th}\frac{df}{dI_\text{dc}}=\frac{\mathcal{N}}{2\pi}\frac{1+Q}{\left ( \zeta +Q \right )^2}.
\end{equation}
The nonlinear frequency shift coefficient $\mathcal{N}$ for STNOs dominates the frequency tunability, and may be positive, zero, or negative, depending on magnetic field direction and magnetic anisotropy of free layer in STNOs.

To explain the experimental observations using this analytical theory, we derive $\mathcal{N}$ with our experimental conditions. The nonlinearity is expressed as~\cite{Slavin2005e} 
\begin{equation}
\label{eq3}
\mathcal{N}=-\frac{\omega _\text{H} \omega _\text{M}\left ( \omega _\text{H}+\omega _\text{M}/4 \right )}{\omega _0\left ( \omega _\text{H}+\omega _\text{M}/2 \right )},
\end{equation}
\begin{equation}
\label{eq4}
\begin{cases}
 & \omega _\text{H}=\gamma H \\ 
 & \omega _\text{M}=4\pi \gamma M_\text{eff} \\ 
 & \omega _0=\gamma \sqrt{\omega _\text{H}\left ( \omega _\text{H}+\omega _\text{M} \right )}.
\end{cases}
\end{equation}

We note that Eqs.~(\ref{eq3}) and (\ref{eq4}) are valid for the magnetization of the free layer being aligned to the magnetic field direction. Utilizing Eqs.~(\ref{eq3}) and (\ref{eq4}), we calculate $df/d\zeta$ ($\propto \mathcal{N}$), where $\zeta $ and Q are used as fitting parameters for all data in Fig.~\ref{fig3}(b), and we find reasonable good agreements with 1.5 for $\zeta $ and 3.0 for Q, respectively. All calculated results are shown as the solid lines alongside the experimental results in Fig.~\ref{fig3}(b). It should be noted that the theoretical calculation coincides with experimental results in the overall trend, although there are discrepancies between experiment and theory. One reason for these discrepancies is likely that the theory does not take into account the current-induced Joule heating and Oersted fields that are present in the experiments. In addition, the calculated nonlinearity $\mathcal{N}$ can also explain the frequency difference between the calculated $f_\mathrm{FMR}$ and the generated microwave frequency $f$ in Fig.~\ref{fig1}. Due to the negative value of $\mathcal{N}$ (or $df/d\zeta$) for NiFe, $f$ is expected to be lower than $f_\mathrm{FMR}$, as predicted in Eq.~(\ref{eq1}) and consistent with our experimental observations in Fig.~\ref{fig1}. 
This auto-oscillation mode is often characterized as a localized bullet \cite{Dumas2013a,Bonetti2010b, Slavin2005e}.
In contrast, $\mathcal{N}$ is positive for easy out-of-plane [Co/Ni], so $f>f_\mathrm{FMR}$ in Fig.~\ref{fig1} \cite{Slavin2005e,Dumas2013a,Chung2018PRL}. In this case, its mode favors to be a propagating spin-wave \cite{Dumas2013a,Madami2015,Mohseni2018b}. All of these experimental observations confirm the theoretical predictions very well.

\begin{figure}
\centering
\includegraphics[width=3.4in]{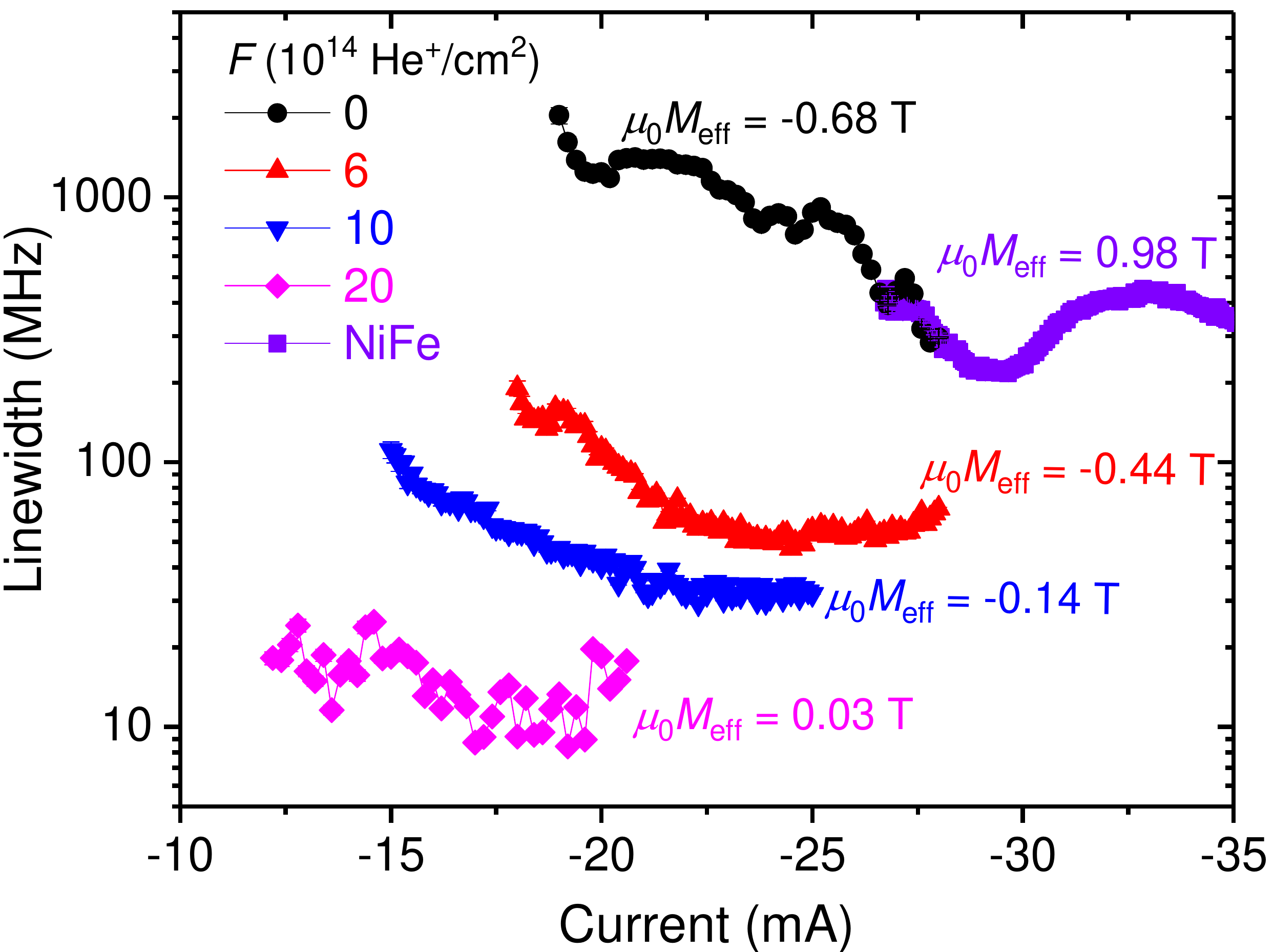}
\caption{\label{fig4} 
Linewidth $\Delta f$ versus $I_\text{dc}$ with different effective magnetization $\mu _0 M_\mathrm{eff}$ at $\mu_0H=0.72$~T. The linewidth were extracted from the data in Figs.~\ref{fig3}(a)--\ref{fig3}(e).
}
\end{figure}

Furthermore, according to nonlinear auto-oscillator theory, the linewidth $\Delta f$ of the generated microwave signals can be expressed as ~\cite{Andrei2009}
\begin{equation}
\label{eq5}
\Delta f=\Gamma _+(P_0)\frac{k_\text{B} T}{E(P_0)}\left [1+\left (\frac{\mathcal{N}} {\Gamma_\text{eff}}  \right )^2\right ],
\end{equation}
where $k_\text{B}$ is the Boltzmann constant and $T$ is the temperature. $\Gamma _+(P_0)$ and $E(P_0)$ are the damping function and time-averaged oscillation energy as a function of the power $P_0$, respectively. $\Gamma _\text{eff}$ is the effective damping. In Eq.(\ref{eq5}), the linewidth $\Delta f$ exhibits a quadratic dependence on the nonlinearity $\mathcal{N}$. To compare with our experimental results, we extracted the linewidth from the data in Figs.~\ref{fig2}(a)--\ref{fig2}(e), as shown in Fig.~\ref{fig4}. The linewidth was indeed dramatically improved by two orders of magnitude as $\mathcal{N}$ decreases (as $M_\mathrm{eff}$ increases), it reaches to a lowest value for $\mu _0 M_\mathrm{eff}=0.03$ T where $\mathcal{N}\rightarrow 0$. $\Delta f$ again increases for the NiFe free layer when $\mathcal{N}$ becomes moderately negative. 
The excellent agreement between our experimental results and theory confirms that the linewidth can be minimized intentionally by controlling the nonlinearity in general, and tuning it to zero in particular. When the PMA compensates the demagnetization field, 
the nonlinearity identically equals zero regardless of the external conditions. 
We can therefore minimize the linewidth by choosing free layer materials with $\mu _0 M_\mathrm{eff}\rightarrow 0$.
We hence would emphasize that our study can offers a  universal path to solving one of the key issues in utilizing STNOs as microwave generators. As for the generated microwave power---another key drawback of this type of microwave generators---we did not observe an improvement in this study, mainly due to the slightly degradation in magnetoresistance (MR) values \cite{Jiang2018a}. We expect that the power can be dramatically improved using magnetic tunnel junction-based STNOs, whose MR can be over two orders of magnitude greater than that of spin valve-based STNOs. \cite{Maehara2013a,Houshang2018b}. 


In conclusion, we have presented a systematic study of the variation of nonlinearity against PMA in STNOs. By using He$^+$ irradiation to continuously tune the PMA of the [Co/Ni] free layer, the nonlinearity $\mathcal{N}$ (along with the frequency tunability $df/dI_\text{dc}$) shows a continuous decreasing trend as $H_\text{k}$ ($M_\text{eff}$) decreases (increases). As a consequence of this decreasing nonlinearity, we have achieved an approximately hundredfold improvement in the linewidth. Our experimental observations are in excellent agreement with nonlinear auto-oscillator theory. This systematic study not only verifies the theoretical prediction, but also offers a route to improving the linewidth, which is of great importance for commercializing microwave generators.

\begin{acknowledgments}
This work was supported by the China Scholarship Council (CSC), the Swedish Foundation for Strategic Research (SSF), the Swedish Research Council (VR), and the Knut and Alice Wallenberg Foundation (KAW). Additional support for the work was provided by  the European Research Council (ERC) under the European Community's Seventh Framework Programme (FP/2007-2013)/ERC Grant 307144 ``MUSTANG''.
\end{acknowledgments}

%

\end{document}